\documentclass[lnicst]{svmultln}
\usepackage{makeidx}  
\usepackage[english]{babel}
\usepackage{graphicx}
\usepackage{amsmath,amssymb}

\newcommand {\ket} [1] {| #1 \rangle}

\newcommand {\cnotcz}  {\textsc{cnot}_{\text{CZ}}}

\newcommand {\mh}  {\hspace{.3 in}}
\newcommand {\sh}  {\hspace{.1 in}}

\newcommand {\bv}  {\bigskip}

\newcommand {\qq} {``}

\begin{document}
\mainmatter              
\title{An Error Model for the Cirac-Zoller \mbox{\sc cnot} gate}
\titlerunning{An Error Model for the Cirac-Zoller \mbox{\sc cnot} gate}
%
\author{Sara Felloni\inst{1,2} \and Giuliano Strini\inst{3}}
\authorrunning{Sara Felloni et al.}   
%
\tocauthor{Sara Felloni, Giuliano Strini}
\institute{Department of Electronics and Telecommunications, Norwegian University of Science and Technology (NTNU), NO-7491 Trondheim, Norway\\
\email{sara.felloni@iet.ntnu.no}
\and
UNIK - University Graduate Center, NO-2027 Kjeller, Norway
\and
Dipartimento di Fisica, Universit\`a degli Studi di Milano,
Via Celoria 16, 20133 Milano, Italy}

\maketitle              

\begin{abstract}
In the framework of ion-trap quantum computing, we develop a characterization of experimentally realistic imperfections which may affect the Cirac-Zoller implementation of the \textsc{cnot} gate.

The \textsc{cnot} operation is performed by applying a protocol of five laser pulses of appropriate frequency and polarization. The laser-pulse protocol exploits auxiliary levels, and its imperfect implementation leads to unitary as well as non-unitary errors affecting the \textsc{cnot} operation.

We provide a characterization of such imperfections, which are physically realistic and have never been considered before to the best of our knowledge. Our characterization shows that imperfect laser pulses unavoidably cause a leak of information from the states which alone should be transformed by the ideal gate, into the ancillary states exploited by the experimental implementation.
\end{abstract}


\section{Decoherence and Quantum Computing by Ion Traps}

Quantum computation and quantum communication require special physical environments: Decoherence, noise and experimental imperfections threaten the correctness of quantum computations, as well as the intended functioning of quantum communication protocols. Several experimental implementations have been proposed and explored in order to build systems capable of having limited sensitivity to unwanted perturbations, while allowing both desired interactions among internal components and access from external systems or users.

Ion-trap quantum computation is one of the main and rapidly evolving experimental possibilities. Several new ideas and experimental techniques \cite{CirZol04} \cite{Iontrap} \cite{HaeRoo08} suggest that ion traps offer a promising architecture for quantum information processing.
In this framework, we develop a characterization of experimentally realistic imperfections which may affect the Cirac-Zoller implementation of the \textsc{cnot} gate \cite{CirZol95}. The \textsc{cnot} operation is performed by applying a protocol of five laser pulses of appropriate frequency and polarization. The laser-pulse protocol exploits auxiliary levels and this could origin unitary as well as non-unitary errors affecting the \textsc{cnot} operation. We provide a characterization of such imperfections, which are physically realistic and have never been considered before to the best of our knowledge.

The paper is organized as follows.
In Section \ref{sec-cz-cnotcz}, we briefly review the sequence of five laser pulses which implement the \textsc{cnot} gate by ion-trap techniques, as proposed by Cirac and Zoller. In Section \ref{sec-cz-noisycz}, we explore the physically realistic perturbations which may affect the laser-pulse protocol for the Cirac-Zoller implementation of the \textsc{cminus} gate and, consequently, of the \textsc{cnot} gate. First, in Section \ref{sec-cz-noisycz-12lev}, we describe by diagrams of quantum states the twelve-level Hilbert space in which we reproduce both the ideal and imperfect actions of the \textsc{cnot} gate. Subsequently, in Section \ref{sec-cz-noisycz-impgt} we model imperfections in all the three laser pulses which implement the \textsc{cminus} gate, by applying perturbations on the parameters determining the physical characteristics of the lasers. Then, in Section \ref{sec-cz-noisycz-kraus} we formulate our error model by means of the well-known density-matrix formalism and Kraus representation, showing how leakage errors are unavoidable in realistic conditions. Finally, in Section \ref{sec-cz-concl} we express our conclusive remarks.

Throughout this paper, we assume the reader is familiar with the basic ion-trap techniques. Useful and detailed physical and computational descriptions of ion-trap quantum computing can be found, for instance, in \cite{BeneCaStqcbook2} (pages 544-561) or \cite{Stea97}.

\section{The Cirac-Zoller \mbox{\sc cnot} Gate}\label{sec-cz-cnotcz}

Ion-trap quantum gates can be obtained by applying to ions implementing qubits the appropriate combination of laser pulses, tuned to the appropriate duration and phase.
In controlled two-qubit operations, the motion state of the string of ions (the \textit{phonon} qubit) is exploited as a ``bus'' to transfer quantum information between two qubits implemented by ions.

Following the proposal of Cirac and Zoller \cite{CirZol95}, we briefly recall how to obtain a \textsc{cnot} quantum gate ($ \cnotcz $) acting on a set of $N$ ions, with the ion $l$ as the control qubit and the ion $m$ as the target qubit.

A general state in the computation is described as $ \ket{i_l , i_m ; i_n}$, where the last position is always filled by the state of the phonon qubit; for the sake of simplicity, we omit all the qubits of the register which are not affected by the protocol. Qubits can be in the ground or excited levels $\ket g$ and $\ket e$; at some point of the protocol, an auxiliary level $\ket a$ is also necessary. The initial state for the computation is $ \ket{i_l , i_m ; 0}$.

First, a red de-tuned laser acts on the ion $l$ in order to map the quantum information of the control ion onto the vibrational mode. This impulse changes the states of the two-qubit computational basis as follows:
\begin{equation}
\left\{
\begin{array}{ccc}
\ket{g_l , g_m; 0} & \rightarrow & \ket{g_l , g_m; 0}\\
\ket{g_l , e_m; 0} & \rightarrow & \ket{g_l , e_m; 0}\\
\ket{e_l , g_m; 0} & \rightarrow & i \ket{g_l , g_m; 1}\\
\ket{e_l , e_m; 0} & \rightarrow & i \ket{g_l , e_m; 1}.\\
\end{array}
\right.
\end{equation}
Then, a red de-tuned laser is applied to the ion $m$.
The impulse is now expressed in the basis $\{ \ket{g_m; n=1}, \ket{a_m; n=0} \}$, thus exploiting the auxiliary level $\ket{a_m}$, and it acts as follows:
\begin{equation}
\left\{
\begin{array}{ccc}
\ket{g_l , g_m; 0} & \rightarrow & \ket{g_l , g_m; 0}\\
\ket{g_l , e_m; 0} & \rightarrow & \ket{g_l , e_m; 0}\\
 i \ket{e_l , g_m; 1} & \rightarrow & - i \ket{e_l , g_m; 1}\\
 i \ket{g_l , e_m; 1} & \rightarrow &  i \ket{g_l , e_m; 1}.\\
\end{array}\right.
\end{equation}
Finally, a red de-tuned laser acts once again on the ion $l$, in order to map the quantum information of vibrational mode back onto the control ion:
\begin{equation}
\left\{
\begin{array}{ccc}
\ket{g_l , g_m; 0} & \rightarrow & \ket{g_l , g_m; 0}\\
\ket{g_l , e_m; 0} & \rightarrow & \ket{g_l , e_m; 0}\\
- i \ket{g_l , g_m; 1} & \rightarrow & \ket{e_l , g_m; 0}\\
 i \ket{g_l , e_m; 1} & \rightarrow & - \ket{e_l , e_m; 0}.\\
\end{array}\right.
\end{equation}

The global effect of the three laser pulses is to induce a controlled phase-shift gate with a phase-shift of an angle $\delta = \pi$, that is, a \textsc{cminus} gate.
From the \textsc{cminus} gate, the \textsc{cnot} gate can be obtained by applying a single-qubit Hadamard gate before and after the three laser pulses.

In conclusion, the $\cnotcz$ operation between the target qubit $m$ and the control qubit $l$ can be performed by applying a protocol of five laser pulses of appropriate frequency and polarization.

\section{Imperfections in the Cirac-Zoller \mbox{\sc cnot} Gate}\label{sec-cz-noisycz}

In the ion-trap implementation of the \textsc{cnot} gate proposed by Cirac and Zoller, a key-role is played by the laser protocol implementing the \textsc{cminus} operation, which exploits \textit{ancillary levels}. The role of such additional levels is here explored to model unitary as well as non-unitary errors affecting the \textsc{cminus} operation and, consequently, the whole implementation of the \textsc{cnot} gate.

The error model schematically consists in the following steps. First, unitary errors are applied to each impulse gate constituting the \textsc{cminus} protocol. Then, a partial trace operation is performed on the collective vibrational motion, whose levels are neglected at the end of both ideal and non-ideal computations. In the ideal protocol, a final step of partial tracing on the ancillary levels would lead to perfectly reproducing the action of the \textsc{cminus} operation. However, in a general experimental situation, our error model shows that it is no longer possible to trace over the ancillary level without loss of information: The unitary errors introduced in the three \textsc{cminus} laser pulses unavoidably cause an irreversible leak of information in the ancillary states.

\subsection{The Hilbert Space}\label{sec-cz-noisycz-12lev}

In order to describe the most general perturbation of the \textsc{cminus} laser-pulse protocol for the Cirac-Zoller implementation of the \textsc{cnot} gate, we rely onto a useful representation and ordering of the twelve-level space which reproduces the action of the gate.

\begin{figure}[!htb]
\begin{center}
\includegraphics[width=6cm]{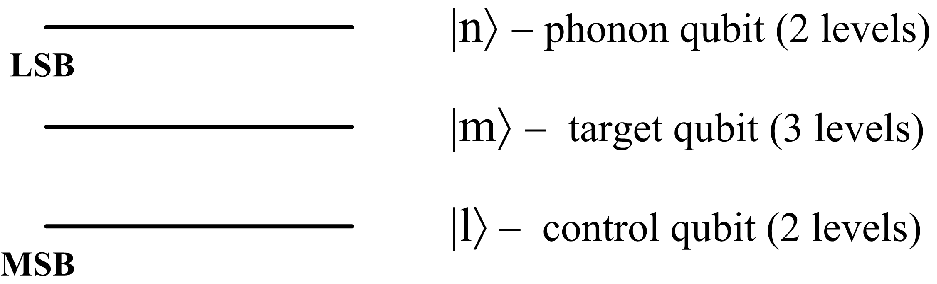}\\
\bv
\includegraphics[width=8cm]{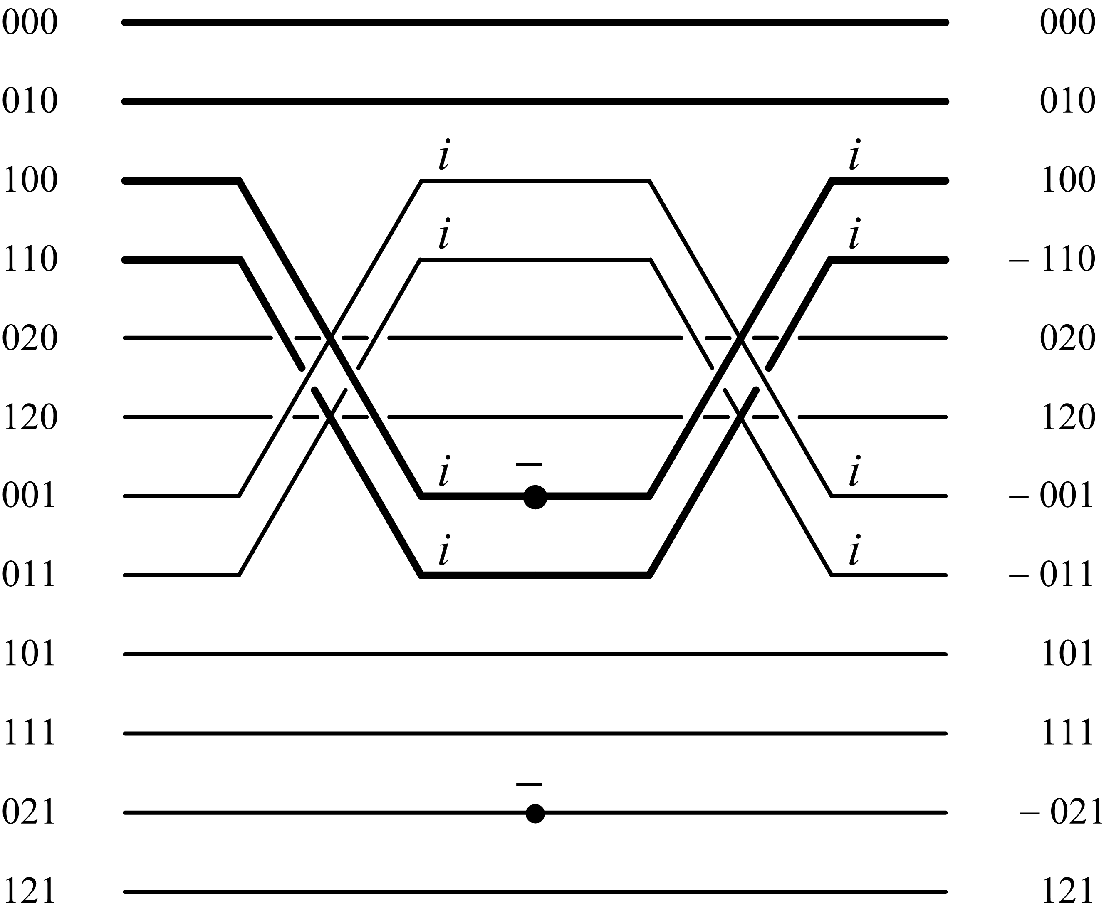}
\end{center}
\caption[Diagram of states \cite{FeStsdI08} representing a noisy ion-trap implementation of the \textsc{cminus} gate, according to the Cirac-Zoller protocol.]{Diagram of states \cite{FeStsdI08} representing a noisy ion-trap implementation of the \textsc{cminus} gate, according to the Cirac-Zoller protocol. The control qubit $m$ in set the most significant position, with possible levels $0$ and $1$; the target qubit $l$ in set the middle, with possible levels $0$, $1$ and $2$ (ancillary); the phonon qubit $n$ in set the least significant position, with possible levels $0$ and $1$. The ordering of the twelve resulting levels is such that the top six levels of the diagram correspond to the non-excited state of the phonon qubit ($n=0$), while the bottom six levels of the diagram correspond to the excited state of the phonon qubit ($n=1$). In each half of the diagram, the two bottom levels correspond to the ancillary state of the target qubit ($m=2$). The sequence of operations must be read from left (input) to right (output); information flows on the thick lines, while thinner lines correspond to
absence of information. Note that the action of the \textsc{cminus} gate can here be visualized by simply following the thick lines corresponding to the first four levels, which alone store all the input information at the start of the gate operation.}\label{cz-diastcminus}
\end{figure}

This ordering is illustrated by the diagram of states in Figure \ref{cz-diastcminus}. In the usual way of representing quantum computations by quantum circuits, each horizontal line represents a qubit. In diagrams of states \cite{FeStsdI08}, we draw instead a horizontal line for each state of the computational basis, here adding horizontal lines for the additional levels necessary to reproduce the action of the \textsc{cnot} gate. Similarly to quantum circuits, any sequence of operations in diagrams of states must be read from left (input) to right (output).

We first represent the ideal action of the three laser pulses, that is, we represent how the information contained in the states of the considered twelve-level space is elaborated when the \textsc{cnot} gate is implemented by a system unaffected by errors or noise. We put the control qubit $m$, which has two possible levels $0$ and $1$, in the most significant position; we put the target qubit $l$, which has two possible standard levels $0$ and $1$ plus an additional ancillary level $2$, in the middle; finally, we put the phonon qubit $n$, which has two possible levels $0$ and $1$, in the least significant position (see upper scheme in Figure \ref{cz-diastcminus}).
We order the resulting twelve levels accordingly (see lower diagram of Figure \ref{cz-diastcminus}). The ordering is such that in the top half of the diagram we have the six levels corresponding to the non-excited state of the phonon qubit, namely, $n=0$; in the bottom half of the diagram we have the remaining six levels, corresponding to the excited state of the phonon qubit, namely, $n=1$. Finally, the two bottom levels of each half of the diagram correspond to the ancillary level of the target qubit, namely, $m=2$.

Since we are considering initial states for which i) the collective vibrational motion is not excited ($n = 0$) and ii) the auxiliary level stores no information, the initial information is stored only in the first four states $\ket{000}$, $\ket{010}$, $\ket{100}$ and $\ket{110}$. The adopted ordering allows us to collect at the top of the diagram the four levels containing the input information, when the collective vibrational motion is not excited and the ancillary states have not yet been exploited. After applying in sequence the three operations corresponding to the three laser pulses previously described, we show at the rightmost end of the diagram how the global transformation affects the twelve levels of the overall system. Again, at the top of the diagram we read on the first four states how the computation reproduces the action of an ideal \textsc{cminus} gate.

\subsection{Imperfect Impulse Gates}\label{sec-cz-noisycz-impgt}

Referring to the Hilbert space previously described, we now model imperfections in the three impulse gates involved in the implementation of the $\cnotcz$ gate.

The action of a general unitary matrix acting on a two-qubit system,
\begin{equation}
U = \left [
\begin{array}{cc}
\cos \frac{\theta}{2} \, ( \cos \psi + i \sin \psi )  &  i \sin \frac{\theta}{2} \, ( \cos \phi + i \sin \phi ) \\
 i \sin \frac{\theta}{2} \, ( \cos \phi - i \sin \phi ) &
\cos \frac{\theta}{2} \, (\cos \psi - i \sin \psi )
\end{array}\right],
\end{equation}
is determined by the three parameters $\theta$, $\psi$, $\phi$. When considering unitary transformations induced by laser pulses, these parameters are associated with precise physical features of the impulses: The parameter $ \theta = \Omega t $ denotes the impulse area, where $ \Omega $ is the intensity and $t$ is the duration of the impulse; the parameter $ \psi $ is related to the laser de-tuning; the parameter $ \phi $ denotes the laser phase.
By appropriately varying these parameters, we obtain the unitary matrices which correspond to the ideal impulses implementing the \textsc{cminus} gate.

The first and the third impulse matrices are obtained by imposing:
\begin{equation}
\{ \theta = \pi, \psi = \pi, \phi = 0 \}, \mh
U^{(1,3)} = \left [
\begin{array}{cc}
0  &  i  \\
i  &  0
\end{array}\right],
\end{equation}
while the second impulse matrix is obtained by imposing:
\begin{equation}
\{ \theta = \pi, \psi = \pi, \phi = 0 \}, \mh
U^{(2)} = \left [
\begin{array}{cc}
- 1  &  0  \\
0  &  - 1
\end{array}\right].
\end{equation}

We apply a perturbation on all the three parameters of all the three impulses, which we will denote from now on with index $\iota$:
\begin{equation}
\{ \theta^\iota + \Delta \theta^\iota, \sh \psi^\iota + \Delta \psi^\iota, \sh \phi^\iota + \Delta \phi^\iota \} \mh \iota=1,2,3.
\end{equation}
Thus, each noisy impulse is now described by:
\begin{equation}
U^{(\iota)} = \left [
\begin{array}{cc}
u^\iota_{11} & u^\iota_{12}\\
u^\iota_{21} & u^\iota_{22}
\end{array}\right]
\end{equation}
where, for $\iota=1,2,3$, we have:
$$
u^\iota_{11} =
\cos \frac{\theta^\iota + \Delta \theta^\iota}{2} ( \cos ( \psi^\iota + \Delta \psi^\iota )  + i \sin ( \psi^\iota + \Delta \psi^\iota ) ),
$$
$$
u^\iota_{12} = i \sin \frac{\theta^\iota + \Delta \theta^\iota}{2} ( \cos ( \phi^\iota + \Delta \phi^\iota ) + i \sin ( \phi^\iota + \Delta \phi^\iota ) ),
$$
$$
u^\iota_{21} = i \sin \frac{\theta^\iota + \Delta \theta^\iota}{2} ( \cos ( \phi^\iota + \Delta \phi^\iota ) - i \sin ( \phi^\iota + \Delta \phi^\iota ) ),
$$
\begin{equation}\label{eq:imppar}
u^\iota_{22} = \cos \frac{\theta^\iota + \Delta \theta^\iota}{2} (\cos ( \psi^\iota + \Delta \psi^\iota ) - i \sin ( \psi^\iota + \Delta \psi^\iota ) ).
\end{equation}

All the three noisy impulse matrices are subsequently embedded in the twelve-level space of the Cirac-Zoller implementation, obtaining respectively the matrices $V^{(1)}$, $V^{(2)}$ and $V^{(3)}$. The entries $\{ u^\iota_{i,j} \}$ are given by the parameters defined by equation (\ref{eq:imppar}); since the matrices are very sparse, we omit all their null entries to obtain a clearer visualization of each matrix structure:
$$
V^{(1)} = \left [
\begin{array}{cccccccccccc}
1 & . & . & . & . & . & . & . & . & . & . & . \\
. & 1 & . & . & . & . & . & . & . & . & . & . \\
. & . & u^1_{11} & . & . & . & u^1_{12} & . & . & . & . & . \\
. & . & . & u^1_{11} & . & . & . & u^1_{12} & . & . & . & . \\
. & . & . & . & 1 & . & . & . & . & . & . & . \\
. & . & . & . & . & 1 & . & . & . & . & . & . \\
. & . & u^1_{21} & . & . & . & u^1_{22} & . & . & . & . & . \\
. & . & . & u^1_{21} & . & . & . & u^1_{22} & . & . & . & . \\
. & . & . & . & . & . & . & . & 1 & . & . & . \\
. & . & . & . & . & . & . & . & . & 1 & . & . \\
. & . & . & . & . & . & . & . & . & . & 1 & . \\
. & . & . & . & . & . & . & . & . & . & . & 1 \\
\end{array}\right]
\sh
V^{(2)} = \left [
\begin{array}{cccccccccccc}
1 & . & . & . & . & . & . & . & . & . & . & . \\
. & 1 & . & . & . & . & . & . & . & . & . & . \\
. & . & 1 & . & . & . & . & . & . & . & . & . \\
. & . & . & 1 & . & . & . & . & . & . & . & . \\
. & . & . & . & 1 & . & . & . & . & . & . & . \\
. & . & . & . & . & 1 & . & . & . & . & . & . \\
. & . & . & . & . & . & u^2_{11} & . & . & . & u^2_{12} & . \\
. & . & . & . & . & . & . & 1 & . & . & . & . \\
. & . & . & . & . & . & . & . & 1 & . & . & . \\
. & . & . & . & . & . & . & . & . & 1 & . & . \\
. & . & . & . & . & . & u^2_{21} & . & . & . & u^2_{22} & . \\
. & . & . & . & . & . & . & . & . & . & . & 1 \\
\end{array}\right]
$$
\begin{equation}
V^{(3)} = \left [
\begin{array}{cccccccccccc}
1 & . & . & . & . & . & . & . & . & . & . & . \\
. & 1 & . & . & . & . & . & . & . & . & . & . \\
. & . & u^3_{11} & . & . & . & u^3_{12} & . & . & . & . & . \\
. & . & . & u^3_{11} & . & . & . & u^3_{12} & . & . & . & . \\
. & . & . & . & 1 & . & . & . & . & . & . & . \\
. & . & . & . & . & 1 & . & . & . & . & . & . \\
. & . & u^3_{21} & . & . & . & u^3_{22} & . & . & . & . & . \\
. & . & . & u^3_{21} & . & . & . & u^3_{22} & . & . & . & . \\
. & . & . & . & . & . & . & . & 1 & . & . & . \\
. & . & . & . & . & . & . & . & . & 1 & . & . \\
. & . & . & . & . & . & . & . & . & . & 1 & . \\
. & . & . & . & . & . & . & . & . & . & . & 1 \\
\end{array}\right].
\end{equation}

Finally, we appropriately embed the Hadamard gates in the twelve-level space of the Cirac-Zoller implementation (still denoting by $H$ the resulting matrices of dimension ${12 \times 12}$, for the sake of simplicity). 

By multiplying the five matrices in the appropriate order, we obtain the overall matrix of dimension ${12 \times 12}$ which describes the action of the $ \cnotcz $ gate:
\begin{equation}\label{eq:Utotal}
    \cnotcz = H \, V^{(3)} \, V^{(2)} \, V^{(1)} \, H.
\end{equation}

\subsection{Density Matrix Evolution and Kraus Operators}\label{sec-cz-noisycz-kraus}

We now calculate the density matrix transformation corresponding to the previously described model for imperfections in the impulse gates originating the $\cnotcz$ gate.

The initial information is stored only in the first four states $\ket{000}$, $\ket{010}$, $\ket{100}$ and $\ket{110}$: This is equivalent to say that all initial information is stored in a density matrix of dimension $ 4 \times 4 $, which we denote by $ \rho^{\, \text{in}}_{\; 4 \times 4}$. Thus, the overall initial density matrix $\rho^{\, \text{in}}_{\; 12 \times 12}$ has non-zero entries only in the ${4 \times 4}$ upper-and-leftmost positions:
\begin{equation}
    \rho^{\, \text{in}}_{\; 6 \times 6} =
\left[
\begin{array}{ccccc}
   & \rho^{\, \text{in}}_{\; 4 \times 4} &  & \mathbf{0}_{\; 4 \times 2} &  \\
   & \mathbf{0}_{\; 2 \times 4} &  & \mathbf{0}_{\; 2 \times 2} &  \\
\end{array}
\right]; \mh
    \rho^{\, \text{in}}_{\; 12 \times 12} =
\left[
\begin{array}{ccccc}
   & \rho^{\, \text{in}}_{\; 6 \times 6} &  & \mathbf{0}_{\; 6 \times 6} &  \\
   & \mathbf{0}_{\; 6 \times 6} &  & \mathbf{0}_{\; 6 \times 6} &  \\
\end{array}
\right].
\end{equation}

The evolution of the overall initial density matrix caused by the laser-pulse protocol is given by:
\begin{equation}\label{eq:rhoevol}
    \rho^{\, \text{fin}}_{\; 12 \times 12} = \cnotcz \; \rho^{\, \text{in}}_{\; 12 \times 12} \; \cnotcz^\dag.
\end{equation}
In order to determine the Kraus operators describing the evolution of the reduced density matrix, the unitary matrix $\cnotcz$ can be decomposed into sub-matrices $A$, $B$, $C$ and $D$, each one of dimension ${6 \times 6}$. Thus, equation (\ref{eq:rhoevol}) can be written as follows:
\begin{equation}
\rho^{\, \text{fin}}_{\; 12 \times 12} = \left[
\begin{array}{ccccc}
   & A &  & B &  \\
   & C &  & D &  \\
\end{array}
\right]
\left[
\begin{array}{ccccc}
   & \rho^{\, \text{in}}_{\; 6 \times 6}  &  & \mathbf{0}_{\; 6 \times 6}  &  \\
   & \mathbf{0}_{\; 6 \times 6}  &  & \mathbf{0}_{\; 6 \times 6}  &  \\
\end{array}
\right]
\left[
\begin{array}{ccccc}
   & A^\dag &  & C^\dag &  \\
   & B^\dag &  & D^\dag &  \\
\end{array}
\right].
\end{equation}

Since at the end of the process implementing the $\cnotcz$ gate we neglect the levels corresponding to the collective vibrational motion, we now perform a partial trace operation on the overall final density matrix $\rho^{ \, \text{fin}}_{ \; 12 \times 12}$, in respect to the phonon qubit $n$:
\begin{equation}
    \rho^{ \, \text{fin}}_{ \; 6 \times 6} = \text{Tr}_{\, \text{n = \textsc{lsb}} \, } \; \{ \, \rho^{ \, \text{fin}}_{ \; 12 \times 12} \, \} = A \; \rho^{\, \text{in}}_{\; 6 \times 6} \, A^\dag + C \, \rho^{\, \text{in}}_{\; 6 \times 6} \, C^\dag.
\end{equation}
We further decompose the sub-matrices $A$ and $C$, both having dimension ${6 \times 6} $, into sub-matrices $\{A_i, C_i\} $, for $i=1 \ldots 4$, each one having the appropriate dimension as expressed in the following equation:
$$
    \rho^{ \, \text{fin}}_{ \; 6 \times 6} = \left[
\begin{array}{ccccc}
   & A1_{\; 4 \times 4} &  & A2_{\; 4 \times 2} &  \\
   & A3_{\; 2 \times 4} &  & A4_{\; 2 \times 2} &  \\
\end{array}
\right]
\left[
\begin{array}{ccccc}
   & \rho^{\, \text{in}}_{\; 4 \times 4} &  & \mathbf{0}_{\; 4 \times 2} &  \\
   & \mathbf{0}_{\; 2 \times 4} &  & \mathbf{0}_{\; 2 \times 2} &  \\
\end{array}
\right]
\left[
\begin{array}{ccccc}
   & A1^{\, \dag}_{\; 4 \times 4} &  & A3^{\, \dag}_{\; 4 \times 2} &  \\
   & A2^{\, \dag}_{\; 2 \times 4} &  & A4^{\, \dag}_{\; 2 \times 2} &  \\
\end{array}
\right] +
$$
\begin{equation}
+
\left[
\begin{array}{ccccc}
   & C1_{\; 4 \times 4} &  & C2_{\; 4 \times 2} &  \\
   & C3_{\; 2 \times 4} &  & C4_{\; 2 \times 2} &  \\
\end{array}
\right]
\left[
\begin{array}{ccccc}
   & \rho^{\, \text{in}}_{\; 4 \times 4} &  & \mathbf{0}_{\; 4 \times 2} &  \\
   & \mathbf{0}_{\; 2 \times 4} &  & \mathbf{0}_{\; 2 \times 2} &  \\
\end{array}
\right]
\left[
\begin{array}{ccccc}
   & C1^{\, \dag}_{\; 4 \times 4} &  & C3^{\, \dag}_{\; 4 \times 2} &  \\
   & C2^{\, \dag}_{\; 2 \times 4} &  & C4^{\, \dag}_{\; 2 \times 2} &  \\
\end{array}
\right].
\end{equation}

Consequently, the final reduced density matrix can be expressed as:
\begin{equation}
    \rho^{\, \text{fin}}_{\; 6 \times 6} =
\left[
\begin{array}{ccccc}
   & \rho1^{\, \text{fin}}_{\; 4 \times 4} &  & \rho2^{\, \text{fin}}_{\; 4 \times 2} &  \\
   & \rho3^{\, \text{fin}}_{\; 2 \times 4} &  & \rho4^{\, \text{fin}}_{\; 2 \times 2} &  \\
\end{array}
\right],
\end{equation}
where:
\begin{equation}
    \rho1^{\, \text{fin}}_{\; 4 \times 4} = A1_{\; 4 \times 4} \; \rho^{\, \text{in}}_{\; 4 \times 4} \; A1^{\, \dag}_{\; 4 \times 4} + C1_{\; 4 \times 4} \; \rho^{\, \text{in}}_{\; 4 \times 4} \; C1^{\, \dag}_{\; 4 \times 4},
\end{equation}
\begin{equation}
    \rho4^{\, \text{fin}}_{\; 2 \times 2} = A3_{\; 2 \times 4} \; \rho^{\, \text{in}}_{\; 4 \times 4} \; A3^{\, \dag}_{\; 4 \times 2} + C3_{\; 2 \times 4} \; \rho^{\, \text{in}}_{\; 4 \times 4} \; C3^{\, \dag}_{\; 4 \times 2}.
\end{equation}

Recalling expression (\ref{eq:Utotal}), we finally explicit the Kraus operators $A1$, $A3$, $C1$ and $C3$:
\begin{equation}
A1_{\; 4 \times 4} = \left [
\begin{array}{cccc}
1 & 0 & 0 & 0 \\
0 & 1 & 0 & 0 \\
0 & 0 & a^1_1 & a^1_2 \\
0 & 0 & a^1_2 & a^1_1 \\
\end{array}\right], \mh
A3_{\; 2 \times 4} = \left [
\begin{array}{cccc}
0 & 0 & 0 & 0 \\
0 & 0 & 0 & 0 \\
\end{array}\right],
\end{equation}
where:
\begin{equation}
a^1_1 = \frac{1}{2} \, (2 u^1_{11} u^3_{11} + u^1_{21} \, (1 + u^2_{11}) \, u^3_{12}), \mh
a^1_2 = \frac{1}{2} \, u^1_{21} \, (- 1 + u^2_{11}) \, u^3_{12},
\end{equation}
with parameters $\{ u^\iota_{i,j} \}$, for $\iota = 1,2,3$, defined by equation (\ref{eq:imppar}), and:
\begin{equation}
C1_{\; 4 \times 4} = \left [
\begin{array}{cccc}
0 & 0 & c^1_1 & c^1_1 \\
0 & 0 & c^1_2 & - c^1_2 \\
0 & 0 & 0 & 0 \\
0 & 0 & 0 & 0 \\
\end{array}\right], \mh
C3_{\; 2 \times 4} = \left [
\begin{array}{cccc}
0 & 0 & c^3 & c^3 \\
0 & 0 & 0 & 0 \\
\end{array}
\right],
\end{equation}
where:
$$
c^1_1 = \frac{1}{\sqrt{2}} \, (u^1_{11} u^3_{21} + u^1_{21} u^2_{11} u^3_{22}),
$$
\begin{equation}
c^1_2 = \frac{1}{\sqrt{2}} \, (u^1_{11} u^3_{21} + u^1_{21} u^3_{22}), \mh
c^3 = \frac{1}{\sqrt{2}} \, u^3_{21} u^2_{21},
\end{equation}
with parameters $\{ u^\iota_{i,j} \}$, for $\iota = 1,2,3$, once again defined by equation (\ref{eq:imppar}).

In conclusion, when the three laser pulses are ideally implemented without perturbation, the overall transformation coincides with the ideal \textsc{cnot} gate, that is, the upper-and-leftmost final density sub-matrix $\rho_{\, \text{main}} = \rho1^{\, \text{fin}}_{\; 4 \times 4}$ is the result of a perfect \textsc{cnot} gate application to the density matrix of the four initial states.
On the other hand, when we deal with inaccurate laser pulses, the final information is no longer stored only in the evolved four initial states. Errors unavoidably cause a spreading of information, which partially remains stored in the four evolved initial states, and partially leaks in the ancillary levels involved in the action of the second laser pulse.

More precisely, the final density matrix previously calculated shows how information is spread by the sequence of imperfect operations. The density sub-matrix $\rho_{\, \text{main}}$ still holds the information corresponding to the imperfect action of the gate on the desired control and target qubits, while the lower-and-rightmost final density sub-matrix $\rho_{\, \text{anc}} = \rho4^{\, \text{fin}}_{\; 2 \times 2}$ holds the information leaked in the auxiliary levels and thus lost from the control and target qubits.

The traces of sub-matrices $\rho_{\, \text{main}}$ and $\rho_{\, \text{anc}}$ give the probabilities $p_{\, \text{main}}$ and $p_{\, \text{anc}}$ that information spreads and populates the main and ancillary levels, respectively. Obviously, $p_{\, \text{main}} + p_{\, \text{anc}} = 1$, as the overall density matrix has unit trace.

\section{Conclusive Remarks}\label{sec-cz-concl}

We have developed a characterization of experimentally realistic imperfections which may affect the Cirac-Zoller implementation of the \textsc{cnot} gate.

In the framework of ion-trap quantum computing, the \textsc{cnot} operation can be performed by applying a protocol of five laser pulses of appropriate frequency and polarization. The laser-pulse protocol exploits auxiliary levels, and this may results in unitary as well as non-unitary errors affecting the \textsc{cnot} operation. We have provided a characterization of such imperfections, which are physically realistic and have never been considered before to the best of our knowledge.
This characterization shows that it is no longer possible to disregard the ancillary levels without loss of information: Unitary errors introduced in the laser pulses unavoidably cause a leak of information in the ancillary states, which can no longer be completely retrieved at the end of the computation.

Modeling experimentally realistic perturbations affecting two-qubit universal operations allow us to explore the impact of two-qubit imperfections on quantum computation and communication protocols. Our previous characterization of general single-qubit errors and its application to a quantum privacy amplification protocol based on entanglement purification \cite{BeneFeSt06} proved that different imperfections can affect a quantum protocol very differently. In a similar way, the modeled two-qubit imperfections may have unexpected effects on the quantum computation and communication protocols simulated in their presence. Extensive simulations will highlight which error parameters are most dangerous and which kinds of perturbations should be expected, especially regarding errors on the ancillary levels.

Finally, by addressing general two-qubit errors affecting universal two-qubit operations, we also aim to extend our previously developed general single-qubit error model.
Since multi-qubit unitary operations can always be computed by equivalent quantum circuits composed of single-qubit and two-qubit gates, a compact model comprising single-qubit and two-qubit errors would offer a satisfiable characterization of general imperfections. Unfortunately, describing and understanding all parameters which characterize the most general transformation of a two-qubit density matrix can not be achieved by straightforward analytical study. Thus, exploring imperfect implementations of universal two-qubit operations in the main current experimental frameworks can offer a valuable help to determine the role and the physical meaning of significant parameters describing the most general two-qubit noise transformation.

\section*{Acknowledgments}

Sara Felloni acknowledges support by ERCIM, as this work was partially carried out during the tenure of an ERCIM \qq Alain Bensoussan'' Fellowship Programme, and she wishes to thank Johannes Skaar for useful insights and discussion.


\begin{thebibliography}{99}
\bibitem{CirZol04} Cirac, J. I., and Zoller, P. (2004), New frontiers in quantum information with atoms and ions, \textit{Phys. Today}, March 2004, 38.
\bibitem{Iontrap} Ion trap approaches to Quantum Information Processing and Quantum Computing - A Quantum Information Science and Technology Roadmap - part 1: Quantum Computation, section 6.2 (2004); \verb"http://qist.lanl.gov/qcomp_map.shtml".
\bibitem{HaeRoo08} Haeffner, H., Roos, C. F., and Blatt, R. (2008), Quantum computing with trapped ions, \textit{Phys. Rep.} \textbf{469}, 155; arXiv:0809.4368.
\bibitem{CirZol95} Cirac, J. I., and Zoller, P. (1995), Quantum computations with cold trapped ions, \textit{Phys. Rev. Lett.} \textbf{74}, 4091.
\bibitem{BeneCaStqcbook2} Benenti, G., Casati, G., and Strini, G. (2007), \textit{Principles of quantum computation and information, Vol. 2: Basic tools and special topics}, World Scientific, Singapore.
\bibitem{Stea97} Steane, A. (1997), The ion trap quantum information processor, \textit{Appl. Phys. B} \textbf{64}, 623.
\bibitem{FeStsdI08} Felloni, S., Leporati, A., and Strini, G. (2008), Diagrams of states in quantum information: An illustrative tutorial, \textit{Int. J. of  Unconv. Comp.}, to appear.
\bibitem{BeneFeSt06} Benenti, G., Felloni, S., and Strini, G. (2006), Effects of single-qubit quantum noise on entanglement purification, \textit{Eur. Phys. J. D} \textbf{38}, 389.
\end{thebibliography}
\end{document}